\documentclass[twocolumn,showpacs,preprintnumbers,amsmath,amssymb,
nofootinbib,superscriptaddress]{revtex4} 

\usepackage{mathrsfs, epsfig, graphicx, url, dcolumn, bm}

\begin{document}

\input{epsf.sty}

\leftline{\hspace{5.9in} UFIFT-QG-08-03}

\title{Gravity Gets There First with Dark Matter Emulators}

\author{S. Desai}
\email{desai@gravity.psu.edu}
\affiliation{Center for Gravitational Wave Physics, Pennsylvania State 
University, University Park, PA 16802}

\author{E. O. Kahya}
\email{emre@phys.ufl.edu}
\author{R. P. Woodard}
\email{woodard@phys.ufl.edu}
\affiliation{Department of Physics, University of Florida, Gainesville,
FL 32611}

\begin{abstract}

We discuss the implications for gravitational wave detectors of a class of
modified gravity theories which dispense with the need for dark matter.
These models, which are known as {\it Dark Matter Emulators}, have the 
property that weak gravitational waves couple to the metric that would
follow from general relativity without dark matter whereas ordinary 
particles couple to a combination of the metric and other fields which
reproduces the result of general relativity with dark matter. We show
that there is an appreciable difference in the Shapiro delays of
gravitational waves and photons or neutrinos from the same source,
with the gravitational waves always arriving first. We compute the expected
time lags for GRB 070201, for SN 1987a, and for Sco-X1. We estimate the
probable error by taking account of the uncertainty in position, and by 
using three different dark matter profiles.

\end{abstract}


\maketitle 

\section{Introduction}
\label{sec:intro}

The direct detection of gravitational waves from astrophysical sources 
would enable us to open a new window into the universe and get insights 
which are complementary to electromagnetic astronomy~\cite{Cutler02}. 
Many ground-based interferometric detectors such as LIGO, VIRGO, GEO600 
and TAMA  have been online for several years. In October 2007, LIGO 
completed a long science run to collect one year of coincident data at
design sensitivity~\cite{LIGO} and the VIRGO detector also joined this 
science run in the last five months. During the latest LIGO science run, 
the sensitivity of the 4~km Hanford and Livingston LIGO detectors to 
detect binary neutron-star coalescence with mass 1.4~$M_{\odot}$ with 
signal to noise ratio greater than 8 (after averaging over all 
orientations and sky positions) was about 15 Mpc~\cite{LIGO}. Analysis 
of the latest LIGO and VIRGO data for gravitational wave (GW) searches 
from a wide variety of sources is in progress~\cite{Papa}.

An important science goal pursued is the search for impulsive transient 
GW signals from sources with electromagnetic and/or neutrino counterparts. 
Some examples of such sources include core-collapse supernovae, gamma-ray 
bursts (GRBs), soft gamma-ray repeaters (SGRs), pulsar glitches, low mass 
X-ray binaries, blazar flares, optical transients, etc~\cite{extrig08}. 
These ``triggered'' searches allow us to get better sensitivity for a given false alarm rate 
as compared to an all-sky search at all times and to design custom-made 
analysis algorithms taking into account our knowledge of the source 
astrophysics. Conversely, there has been a proposal to look for optical 
and infrared counterparts at the time of coincident GW burst 
candidates~\cite{Kanner08}. An overview and benefits of such triggered 
searches carried out by the current interferometric gravitational wave 
detectors are reviewed in Ref.~\cite{extrig08}. There have been proposals 
to determine neutrino mass using simultaneous neutrino and GW observations 
from core-collapse supernova~\cite{Arnaud01}. Similar triggered GW  
searches will also be important for the future LISA experiment~\cite{Kocsis}.

In all present and past triggered searches for gravitational waves, the 
analysis is done  by looking at the data from GW detectors within a 
narrow time window (of about hundreds of seconds) around the time of 
the electromagnetic trigger. With this assumption, one can detect 
gravitational waves  only if the propagation time of photons/neutrinos 
is the same as that of gravitational waves. In general relativity, photons, 
neutrinos and gravitational waves propagate on the same null geodesics. 
Hence the total time of propagation is the light travel time delay plus 
the Shapiro time-delay due to intervening matter~\cite{Shapiro64}. For 
electromagnetic waves, Shapiro delay has been detected in a wide 
variety of systems such as radar ranging to Venus, Doppler tracking 
of Cassini spacecraft and in binary pulsars~\cite{Will06}.
From the relative arrival times of photons and neutrinos from SN 1987a, 
we also know that the Shapiro time-delay for neutrinos is the same as that 
for photons to within 0.5\%~\cite{Longo88,Krauss88}.

The conventional view is that general relativity describes gravity on
cosmic scales. If this is so, the gravitation of stars and gas is not
sufficient to account for the velocity dispersions in clusters~\cite{Zwi},
or for the rotation curves of spiral galaxies~\cite{RTF,RFT1,RFT2}, or for 
the weak lensing in galactic clusters~\cite{TVW,FKSW,SKFBW,CLKHG,Mellier,WTKAB}.
Big Bang Nucleosynthesis severely limits the extent to which the deficit
can be made up of unseen but ordinary matter~\cite{Olive}; the remainder
must consist of an exotic, nonrelativistic substance which has never been 
detected except gravitationally. This {\it dark matter} must vastly
predominate over ordinary matter. For example, only about one fifth of our 
galaxy's mass is made up of normal matter, with the rest being composed of 
dark matter~\cite{Trimble87}. Thus the dominant contribution to the Shapiro
delay for photons from GRBs and other sources is due to the gravitational 
potential of the intervening dark matter.

None of the proposed dark matter candidates have been detected, either 
directly in a laboratory experiment or indirectly through their annihilation 
products~\cite{Bertone04,ADMX,Hooper08,CDMS}. This prompts the suspicion that
perhaps it is gravity which must be modified, rather than the universe's
inventory of nonrelativistic matter. Of course that would invalidate the
assumption which is the basis for all current and proposed GW searches 
from sources seen in photons and neutrinos~\cite{extrig08,Arnaud01}. A 
previous study~\cite{Kahya07,Kahya08} has considered the consequences for 
gravitational wave detection of a certain class of modified gravity theories 
known as {\it Dark Matter Emulators}. In this paper we correct a mistake 
in the original work~\cite{Kahya07} that led to the wrong sign for the 
effect, and we work out explicit results for three interesting sources.

Section II defines and motivates Dark Matter Emulators. In Section III we 
review three popular dark matter profiles which these models are designed 
to obviate. Section IV computes the expected time lag between the arrival 
of the pulse of gravitational waves from some cosmic event and its optical or 
neutrino counterpart. In Section V we give explicit results for three 
sources of interest. Section VI gives a very brief discussion of other 
alternate gravity models, and our conclusions comprise Section VII.

\section{Dark Matter Emulators}

Certain regularities in cosmic structures suggest modified gravity.
One of these is the Tully-Fisher relation, which states that the
luminosity of a spiral galaxy is proportional to the fourth power
of the peak velocity in its rotation curve \cite{TF}. If luminous
matter is insignificant compared to dark matter, why should such a
relation exist? Another regularity is Milgrom's Law, which states
that the need for dark matter occurs at gravitational accelerations
of $a_0 \simeq 10^{-10}~{\rm m/s}^2$ \cite{KT}. A third regularity
is that $a_0$ also seems to give the internal accelerations of 
pressure-supported objects ranging over six orders of magnitude in 
size --- from massive molecular clouds within our own galaxy to 
X-ray clusters of galaxies \cite{SM}.

A modification of Newtonian gravity which explains these regularities
was proposed by Milgrom in 1983 \cite{Milg}. His model, Modified
Newtonian Dynamics (MOND), was soon given a Lagrangian formulation in
which conservation of energy, 3-momentum and angular momentum are
manifest \cite{BM}. However, there was for years no successful
relativistic generalization which could be employed to study
cosmological evolution. Even in the context of static, spherically
symmetric geometries,
\begin{equation}
ds^2 \equiv -B(r) c^2 dt^2 + A(r) dr^2 + r^2 d\Omega^2 \; , \label{ds2}
\end{equation}
the early formulation of MOND fixed only $B(r)$, not $A(r)$. It was
therefore incapable of making definitive predictions about gravitational
lensing.

A relativistic extension of MOND has recently been proposed by Bekenstein
\cite{Bek}. This model is known as TeVeS for ``Tensor-Vector-Scalar.''
In addition to reproducing the MOND force law at low accelerations,
TeVeS has acceptable post Newtonian parameters, and it gives a
plausible amount of gravitational lensing \cite{Bek}. When TeVeS is
used in place of general relativity + dark matter to study cosmological
evolution, the results are in better agreement with data than many
thought possible \cite{Skordis,Skordis2,Dodelson,Bourliot,ZFS}. The model
does have problems with stability~\cite{Clayton,CWW}. The Bullet Cluster
is sometimes cited as a fatal blow for the model~\cite{SMC} but opinion 
on this differs~\cite{AFZ,Stacy,FFB}, and this system in any case poses 
problems for dark matter~\cite{BrM2,Stacy}.

What concerns us here is the curious property of TeVeS that small amplitude
gravitational waves are governed, as in  general relativity, by the metric
$g_{\mu\nu}$, whereas matter couples to a ``disformally transformed'' metric
which involves the vector and scalar fields,
\begin{equation}
\widetilde{g}_{\mu\nu} = e^{-2\phi} (g_{\mu\nu} + A_{\mu} A_{\nu})
- e^{2\phi} A_{\mu} A_{\nu} \; .
\end{equation}
The Scalar-Vector-Tensor gravity (SVTG) theory proposed by Moffat also
has different metrics for matter and small amplitude gravitational waves
\cite{JWM,BrM}. The appearance of this feature in two very different
models is the result of trying to reconcile solar system tests with modified
gravity at ultra-low accelerations. Solar system tests strongly predispose
the Lagrangian to possess an Einstein-Hilbert term \cite{Will,BEF}. On the 
other hand, failed attempts to generalize MOND \cite{SW1} have led to a 
theorem that one cannot get sufficient weak lensing from a stable, covariant 
and purely metric theory which reproduces the Tully-Fisher relation without
dark matter \cite{SW2}. Hence the MOND force must be carried by some other
field, and it is a combination of this other field and the metric which
determines the geodesics for ordinary matter. However, the dynamics of
small amplitude gravitational waves are still set by the linearized Einstein
equation. This simple observation makes for a sensitive and generic test.

We define a {\it Dark Matter Emulator} as any modified gravity theory for
which:
\begin{enumerate}
\item{Ordinary matter couples to the metric $\widetilde{g}_{\mu\nu}$
that would be produced by general relativity + dark matter; and}
\item{Small amplitude gravitational waves couple to the metric $g_{\mu\nu}$
produced by general relativity without dark matter.}
\end{enumerate}
Now consider a cosmic event such as a supernova which emits simultaneous
pulses of gravitational waves and either neutrinos or photons. If physics is
described by a dark matter emulator then the pulse of gravitational waves will
reach us on a lightlike geodesic of $g_{\mu\nu}$, whereas neutrinos and
photons travel along a lightlike geodesic of $\widetilde{g}_{\mu\nu}$.
If significant propagation occurs over regions that would be dark matter
dominated in general relativity then there will be a measurable lag
between arrival times. 

Currently the only observational constraint on the speed $v_{\rm g}$ of
gravity relative to that of ordinary matter $v_{\rm m}$ derives from the 
consequences of gravitational Cherenkov radiation from particles moving 
faster than gravity~\cite{Caves,Moore01}. From observations of the 
highest energy cosmic rays Moore and Nelson infer the bound, $v_{\rm m} 
- v_{\rm g} < 2 \times 10^{-15} c$~\cite{Moore01}. Although the original 
study of Dark Matter Emulators~\cite{Kahya07} in fact violated this bound, 
that was the result of incorrectly choosing the dimensional constant in 
a certain logarithm. In the next section we show that the speed of gravity 
is always greater than that of light for Dark Matter Emulators. A discussion of the Shapiro delay 
calculation in some other  alternate gravity theories can be found in Refs.~\cite{Carlip04,Asada07}.

\section{Three Dark Matter Profiles}

We shall specialize to static, spherically symmetric distributions of dark 
matter, consistent with the invariant element (\ref{ds2}). It is well to
bear in mind that hierarchical structure formation will not necessarily 
result in spherically symmetric distributions~\cite{WNEF}. There is even 
evidence that the dark matter halo conjectured to surround our own galaxy 
is not spherical~\cite{LJM}.

For a pressureless, static, spherically symmetric system the Einstein 
equations take the form,
\begin{eqnarray}
\frac{B}{A} \Biggl[ \frac{A'}{r A} + \Bigl(\frac{A-1}{r^2}\Bigr)\Biggr]
& = & \frac{8\pi G}{c^2} \, \rho \; , \label{Eqn1} \\
\frac{B'}{r B} - \Bigl(\frac{A-1}{r^2}\Bigr) & = & 0 \; . \label{Eqn2}
\end{eqnarray}
If the potential $B(r)$ goes to a constant at infinity we can choose the
time units so that equation (\ref{Eqn2}) has the exact solution,
\begin{equation}
B(r) = \exp\Biggl[-\int_r^{\infty} dr' \, \Bigl(\frac{A(r') - 1}{r'}\Bigr)
\Biggr] \; .
\end{equation}
However, our study requires only small corrections to $A(r)$ and $B(r)$,
\begin{equation}
A(r) \equiv 1 + \Delta A(r) \qquad , \qquad B(r) \equiv 1 + \Delta B(r) \; .
\end{equation}
This not only simplifies (\ref{Eqn1}-\ref{Eqn2}), it also means we can 
dispense with the contribution of ordinary matter to the mass density 
$\rho(r)$. The reason is that we are computing the difference in propagation
times along null geodesics between the same points in two different 
geometries. The geometry felt by gravitational waves is sourced only by ordinary
matter, while the geometry felt by photons and neutrinos is sourced by the
sum of ordinary matter and dark matter. At first order in the mass density, 
the effect of ordinary matter cancels out when computing the difference in 
propagation times between the two geometries! We will henceforth consider
$\rho(r)$ to be the density of dark matter.

The potentials $\Delta A(r)$ and $\Delta B(r)$ can be given simple 
expressions in terms of the mass function,
\begin{equation}
M(r) \equiv 4 \pi \int_0^r dr' \, \rho(r') \; .
\end{equation}
The linearized solution of (\ref{Eqn1}) is,
\begin{equation}
\Delta A(r) = \frac{8 \pi G}{c^2 r} \int_0^r dr' \, r^{\prime 2} \rho(r') =
\frac{2 G}{c^2} \, \frac{M(r)}{r} \; . \label{DA}
\end{equation}
Note that $\Delta A(r)$ is positive semi-definite. From (\ref{DA})
we find $\Delta B(r)$,
\begin{equation}
\Delta B(r) = -\!\!\int_r^{\infty} \!\!\!\!\! dr' \, \frac{\Delta A(r')}{r'} = 
-\Delta A(r) - \frac{2 G}{c^2} \!\!\!\int_r^{\infty} \!\!\!\!\! dr' \, 
\frac{M(r')}{r'} \; . \label{DBfromDA}
\end{equation}
Note that $\Delta B(r)$ is negative semi-definite and in fact less than or 
equal to $-\Delta A(r)$. This guarantees that gravitational waves travel faster 
than photons or neutrinos, so there is no problem with the bound of Moore 
and Nelson~\cite{Moore01}.

Our study requires the dark matter density functions for our own galaxy and 
(for the most distant source) for the Andromeda galaxy. We took these in 
the form of fits to three popular density profiles whose analytic forms
are presented at the end of this section. Given the current rough quality
of the observational data, a Dark Matter Emulator that reproduced the
potentials $\Delta A(r)$ and $\Delta B(r)$ for any of these profiles would 
be judged successful. One can therefore regard the slightly different time
delays that result as one measure of the theoretical uncertainty. The fits 
for our own galaxy appear in Table~\ref{MWPs} and were taken from the study by 
Ascasibar, Jean, Boehm and Kn\"odlseder of the positron annihilation line 
from the galactic center~\cite{Ascasibar}. The fits for Andromeda appear in 
Table~\ref{M31Ps} and were done by Tempel, Tam and Tenjes~\cite{Tempel}.

\begin{table}

\vbox{\tabskip=0pt \offinterlineskip
\def\tablerule{\noalign{\hrule}}
\halign to245pt {\strut#& \vrule#\tabskip=1em plus2em& \hfil#\hfil&
\vrule#& \hfil#\hfil& 
\vrule#& \hfil#\hfil& \vrule#& \hfil#\hfil& \vrule#\tabskip=0pt\cr
\tablerule 
\omit&height4pt&\omit&&\omit&&\omit&&\omit&\cr
&&$\!\!\!\!{\rm Profile}\!\!\!\!$ && $\!\!\!\!\! 8 \pi G \rho_0 r_0^3/c^3
\!\!\!\!\!$ && $\!\!\!\!\! r_0 \!\!\!\!\!$ && $\!\!\!\!\! r_c \!\!\!\!\!$ & \cr 
\omit&height4pt&\omit&&\omit&&\omit&&\omit&\cr
\tablerule 
\omit&height4pt&\omit&&\omit&&\omit&&\omit&\cr
&& $\!\!\!\!{\rm Isothermal}\!\!\!\!$ && $\!\!\!\!\! 3.98~{\rm days} 
\!\!\!\!\!$ && $\!\!\!\!\! 4.00~{\rm kpc} \!\!\!\!\!$ 
&& $\!\!\!\!\! 219~{\rm kpc} \!\!\!\!\!$ & \cr 
\omit&height4pt&\omit&&\omit&&\omit&&\omit&\cr
\tablerule
\omit&height4pt&\omit&&\omit&&\omit&&\omit&\cr
&&$\!\!\!\!{\rm NFW}\!\!\!\!$ && $\!\!\!\!\! 60.8~{\rm days} \!\!\!\!\!$ && 
$\!\!\!\!\! 16.7~{\rm kpc} \!\!\!\!\!$ && $\!\!\!\!\! {\rm N.A.} 
\!\!\!\!\!$ & \cr 
\omit&height4pt&\omit&&\omit&&\omit&&\omit&\cr
\tablerule
\omit&height4pt&\omit&&\omit&&\omit&&\omit&\cr
&& $\!\!\!\!{\rm Moore}\!\!\!\!$ && $\!\!\!\!\! 51.8~{\rm days} \!\!\!\!\!$ && 
$\!\!\!\!\! 29.5~{\rm kpc} \!\!\!\!\!$ && $\!\!\!\!\! {\rm N.A.} 
\!\!\!\!\!$ & \cr 
\omit&height4pt&\omit&&\omit&&\omit&&\omit&\cr
\tablerule}}

\caption{Dark Matter Profile Parameters for the Milky Way Galaxy from
Ascasibar {\it et al.}~\cite{Ascasibar}.}
\label{MWPs}
\end{table}

\begin{table}

\vbox{\tabskip=0pt \offinterlineskip
\def\tablerule{\noalign{\hrule}}
\halign to245pt {\strut#& \vrule#\tabskip=1em plus2em& \hfil#\hfil&
\vrule#& \hfil#\hfil& 
\vrule#& \hfil#\hfil& \vrule#& \hfil#\hfil& \vrule#\tabskip=0pt\cr
\tablerule 
\omit&height4pt&\omit&&\omit&&\omit&&\omit&\cr
&&$\!\!\!\!{\rm Profile}\!\!\!\!$ && $\!\!\!\!\! 8 \pi G \rho_0 r_0^3/c^3
\!\!\!\!\!$ && $\!\!\!\!\! r_0 \!\!\!\!\!$ && $\!\!\!\!\! r_c \!\!\!\!\!$ & \cr 
\omit&height4pt&\omit&&\omit&&\omit&&\omit&\cr
\tablerule 
\omit&height4pt&\omit&&\omit&&\omit&&\omit&\cr
&& $\!\!\!\!{\rm Isothermal}\!\!\!\!$ && $\!\!\!\!\! 1.88~{\rm days} 
\!\!\!\!\!$ && $\!\!\!\!\! 1.47~{\rm kpc} \!\!\!\!\!$ 
&& $\!\!\!\!\! 117~{\rm kpc} \!\!\!\!\!$ & \cr 
\omit&height4pt&\omit&&\omit&&\omit&&\omit&\cr
\tablerule
\omit&height4pt&\omit&&\omit&&\omit&&\omit&\cr
&&$\!\!\!\!{\rm NFW}\!\!\!\!$ && $\!\!\!\!\! 48.6~{\rm days} \!\!\!\!\!$ && 
$\!\!\!\!\! 12.5~{\rm kpc} \!\!\!\!\!$ && $\!\!\!\!\! {\rm N.A.} 
\!\!\!\!\!$ & \cr 
\omit&height4pt&\omit&&\omit&&\omit&&\omit&\cr
\tablerule
\omit&height4pt&\omit&&\omit&&\omit&&\omit&\cr
&& $\!\!\!\!{\rm Moore}\!\!\!\!$ && $\!\!\!\!\! 45.8~{\rm days} \!\!\!\!\!$ && 
$\!\!\!\!\! 25.0~{\rm kpc} \!\!\!\!\!$ && $\!\!\!\!\! {\rm N.A.} 
\!\!\!\!\!$ & \cr 
\omit&height4pt&\omit&&\omit&&\omit&&\omit&\cr
\tablerule}}

\caption{Dark Matter Profile Parameters for the Andromeda Galaxy from
Tempel {\it et al.}~\cite{Tempel}.}
\label{M31Ps}
\end{table}

\subsection{Isothermal Halo Profile}

We shall use a variant of the Isothermal Halo Profile in which the density
vanishes after a cutoff radius $r_c$~\cite{King,Einasto},
\begin{equation}
\rho(r) = \Biggl[ \frac{\rho_0}{1 + (\frac{r}{r_0})^2} - 
\frac{\rho_0}{1 + (\frac{r_c}{r_0})^2} \Biggr] \theta(r_c - r) \; .
\end{equation}
Such a cutoff is inevitable, even in MOND, owing to the presence of other
galaxies. Of course it is also necessary to make the potential $\Delta B(r)$
vanish at infinity.

For $r < r_c$ the mass function and potentials are,
\begin{eqnarray}
M(r) & \!\!\!=\!\!\! & 4\pi \rho_0 r_0^3 \Biggl\{\frac{r}{r_0}
\!-\! \tan^{-1}\Bigl(\frac{r}{r_0}\Bigr) \!-\!\frac{r^3}{3 r_0 (r_0^2 + r_c^2)} 
\Biggr\} , \qquad \\
\Delta A(r) & \!\!\! =\!\!\! & \frac{8\pi G \rho_0 r_0^2}{c^2}
\Biggl\{ \!\! 1 \!-\! \frac{r_0}{r} \tan^{-1}\!\Bigl(\frac{r}{r_0}\Bigr) \!-\!
\frac{r^2}{3 (r_0^2 + r_c^2)} \!\Biggr\} , \qquad \\
\Delta B(r) & \!\!\!= \!\!\!& \frac{8\pi G \rho_0 r_0^2}{c^2}
\Biggl\{ -1 \!+\! \frac{r_0}{r} \tan^{-1}\Bigl(\frac{r}{r_0}\Bigr) \nonumber \\
& & \hspace{1.7cm} + \frac{3 r_c^2 - r^2}{6 (r_0^2 + r_c^2)} - \frac12 
\ln\Bigl[\frac{r_c^2 + r_0^2}{r^2 + r_0^2}\Bigr] \Biggr\} . \qquad
\end{eqnarray}
For $r > r_c$ the mass is constant and the (equal and opposite) potentials 
fall off like $1/r$,
\begin{eqnarray}
M(r) & \!\!\!=\!\!\! & 4\pi \rho_0 r_0^3 \Biggl\{\frac{r_c}{r_0}
\!-\! \tan^{-1}\Bigl(\frac{r_c}{r_0}\Bigr) \!-\!\frac{r_c^3}{3 r_0 (r_0^2 
+ r_c^2)} \Biggr\} , \qquad \\
\Delta A(r) & \!\!\! =\!\!\! & \frac{8\pi G \rho_0 r_0^2}{c^2} \Biggl\{\!\!
\frac{r_c (2 r_c^2 + 3 r_0^2)}{3 r (r_c^2 + r_0^2)} - \frac{r_0}{r} 
\tan^{-1}\!\Bigl(\frac{r_c}{r_0}\Bigr) \!\!\Biggr\} , \qquad \\
\Delta B(r) & \!\!\! =\!\!\! & \frac{8\pi G \rho_0 r_0^2}{c^2} \Biggl\{\! -
\frac{r_c (2 r_c^2 \!+\! 3 r_0^2)}{3 r (r_c^2 \!+\! r_0^2)} \!+\! \frac{r_0}{r} 
\tan^{-1}\!\Bigl(\frac{r_c}{r_0}\Bigr) \!\!\Biggr\} . \qquad
\end{eqnarray}

It should be noted that Ascasibar, Jean, Boehm and Kn\"odlseder used a
simpler version of the isothermal profile without a cutoff radius $r_c$.
This would cause the potential $\Delta B(r)$ to eventually become 
positive, which violates the bound of Moore and Nelson~\cite{Moore01}.
It also doesn't make any sense when one considers the effect of other
galaxies. Because the isothermal profile is the most closely related to
MOND we considered it important to include results for this profile, so 
we used the values of $\rho_0$ and $r_0$ given by Ascasibar, Jean, Boehm 
and Kn\"odlseder~\cite{Ascasibar}, along with $r_c = 219~{\rm kpc}$. 
This choice for $r_c$ causes the ratio of the total masses of the Milky 
Way and Andromeda galaxies for the isothermal profile to agree with that
of the NFW profile considered in the next subsection.

\subsection{NFW Profile}

The NFW profile was the result of studying the equilibrium density
profiles of dark matter halos in numerical simulations of structure
formation~\cite{NFW},
\begin{equation}
\rho(r) = \frac{\rho_0}{\frac{r}{r_0} [1 + \frac{r}{r_0}]^2} \; .
\end{equation}
The associated mass function and potentials are,
\begin{eqnarray}
M(r) & = & 4\pi \rho_0 r_0^3 \times \Biggl\{ \ln\Bigl[1 + \frac{r}{r_0}\Bigr]
- \frac{r}{r_0 + r}\Biggr\} \; , \\
\Delta A(r) & = & \frac{8\pi G \rho_0 r_0^2}{c^2} \times \Biggl\{ \frac{r_0}{r}
\ln\Bigl[1 + \frac{r}{r_0}\Bigr] - \frac{r_0}{r_0 + r} \Biggr\} \; , \qquad \\
\Delta B(r) & = & \frac{8\pi G \rho_0 r_0^2}{c^2} \times -\frac{r_0}{r}
\ln\Bigl[1 + \frac{r}{r_0}\Bigr] \; .
\end{eqnarray}

\subsection{Moore Profile}

A later effort along the same lines showed a better fit to a density
function which is more sharply peaked at the center~\cite{Moore},
\begin{equation}
\rho(r) = \frac{\rho_0}{(\frac{r}{r_0})^{\frac32} [1 + 
\frac{r}{r_0}]^{\frac32}} \; .
\end{equation}
The associated mass function and potentials are,
\begin{eqnarray}
M(r) & = & 4\pi \rho_0 r_0^3 \times \frac23 \ln\Biggl[1 + 
\Bigl(\frac{r}{r_0}\Bigr)^{\frac32}\Biggr] \; , \\
\Delta A(r) & = & \frac{8\pi G \rho_0 r_0^2}{c^2} \times \frac23 \frac{r_0}{r}
\ln\Biggl[1 + \Bigl(\frac{r}{r_0}\Bigr)^{\frac32}\Biggr] \; , \\
\Delta B(r) & = & \frac{8\pi G \rho_0 r_0^2}{c^2} \times \Biggl\{
-\frac23 \frac{r_0}{r} \ln\Biggl[1 + \Bigl(\frac{r}{r_0}\Bigr)^{\frac32}
\Biggr] \nonumber \\
& & + \ln\Biggl[1 + \Bigl(\frac{r_0}{r}\Bigr)^{\frac12}\Biggr] 
-\frac13 \ln\Biggl[1 + \Bigl(\frac{r_0}{r}\Bigr)^{\frac32}\Biggr] \nonumber \\
& & \hspace{1.8cm} - \frac2{\sqrt{3}} \tan^{-1}\Biggl[\frac{\sqrt{3 r_0}}{2 
\sqrt{r} - \sqrt{r_0}} \Biggr] \Biggr\} \; . \qquad
\end{eqnarray}

\section{The Shapiro Delay for Given $M(r)$}

It is more convenient to convert the spatial coordinates from spherical 
$(r,\theta,\phi)$ to Cartesian $x^i$,
\begin{equation}
\vec{x} \equiv r \Bigl( \sin\theta \cos\phi, \sin\theta \sin\phi, 
\cos\theta \Bigr) \equiv r \hat{r} \; .
\end{equation}
The invariant element (\ref{ds2}) has a simple expression in terms of
these coordinates,
\begin{equation}
ds^2 = -c^2 dt^2 + d\vec{x} \cdot d\vec{x} -\Delta B \, c^2 dt^2 + 
\Delta A \, (\hat{r} \cdot d\vec{x})^2 \; . \label{line1}
\end{equation}
Before moving on we digress to note that specializing to $ds^2 = 0$ and 
identifying the velocity of photons and neutrinos as $\vec{v}_{\rm m} = 
d\vec{x}/dt$ results in an equation for the speed of effectively massless,
ordinary matter,
\begin{equation}
0 = -(1 + \Delta B) c^2 + \vec{v}_{\rm m} \cdot \vec{v}_{\rm m} +
\Delta A (\hat{r} \cdot \vec{v}_{\rm m})^2 \; . \label{veq}
\end{equation}
Now recall that we are ignoring the role of ordinary matter in both 
the metrics of gravity (where it is the only part of the mass density) 
and of ordinary matter. It follows that the speed of gravity is $v_{\rm g}$
is $c$. Treating to first order in the potentials and assuming $\Delta A(r) 
\geq 0$ and $\Delta B(r) \leq 0$, we see that $v_{\rm m} - v_{\rm g} \leq 0$.

We can express the invariant element (\ref{line1}) as the flat space 
contribution plus a perturbation,
\begin{equation}
ds^2 \equiv (\eta_{\mu\nu} + h_{\mu\nu}) dx^{\mu} dx^{\nu} \; . \label{line2}
\end{equation}
Comparing (\ref{line1}) and (\ref{line2}) allows us to read off the 
$3+1$ decomposition of the graviton field $h_{\mu\nu}$,
\begin{equation}
h_{00} = -\Delta B \quad , \quad h_{0i} = 0 \quad {\rm and} \quad h_{ij} =
\Delta A \, \hat{r}^i \hat{r}^j \; .
\end{equation}
One advantage of Cartesian coordinates is that the affine connection
vanishes for the flat background. It is easy to give the first correction,
\begin{eqnarray}
\Gamma^{\mu}_{~\rho\sigma} & = & \Delta \Gamma^{\mu}_{~\rho\sigma} 
+ O(h^2) \; , \\
\Delta \Gamma^{\mu}_{~\rho\sigma} & = & \frac12 \eta^{\mu\nu} \Bigl(
h_{\nu \rho , \sigma} + h_{\sigma \nu , \rho} - h_{\rho \sigma , \nu}
\Bigr) \; .
\end{eqnarray}

We need the null geodesic $\chi^{\mu}(\tau)$ that connects the spacetime 
points $x_1^{\mu} = (0,\vec{x}_1)$ and $x_2^{\mu} = (ct,\vec{x}_2)$. It
obeys the geodesic equation,
\begin{equation}
\ddot{\chi}^{\mu} + \Gamma^{\mu}_{~\rho\sigma}\Bigl(\chi(\tau)\Bigr)
\dot{\chi}^{\rho} \dot{\chi}^{\sigma} = 0 \; ,
\end{equation}
subject to the conditions,
\begin{eqnarray}
\chi^{\mu}(0) & = & x_1^{\mu} \; , \\
\chi^i(1) & = & x_2^i \; , \\
g_{\mu\nu}(x_1) \dot{\chi}^{\mu}(0) \dot{\chi}^{\nu}(0) & = & 0 \; .
\end{eqnarray}
Of course we solve this perturbatively in the potentials. The zeroth
order solution is of course the flat space result,
\begin{equation}
\chi_0^{\mu}(\tau) = x_1^{\mu} + \Delta x^{\mu} \tau \; .
\end{equation}
Here the temporal and spatial components of the interval $\Delta x^{\mu}$
are,
\begin{equation}
\Delta x^0 \equiv \Vert \vec{x}_2 - \vec{x}_1\Vert \qquad {\rm and} \qquad
\Delta x^i \equiv x_2^i - x_1^i \; .
\end{equation}

The first order corrections to the spatial components of the geodesic are,
\begin{eqnarray}
\lefteqn{\chi_1^i(\tau) = \tau \int_0^1 \! d\tau' \, (1-\tau') 
\Delta \Gamma^i_{~\rho\sigma}\Bigl(x + \Delta x \tau\Bigr) \Delta x^{\rho}
\Delta x^{\sigma} } \nonumber \\
& & \hspace{.5cm} - \int_0^{\tau} \! d\tau' \, (\tau - \tau') 
\Delta \Gamma^i_{~\rho\sigma}\Bigl(x_1 + \Delta x \tau\Bigr) \Delta x^{\rho}
\Delta x^{\sigma} \; . \qquad
\end{eqnarray}
Of course it is from the first order temporal correction that we infer
the time lag. This correction is more complicated,
\begin{eqnarray}
\lefteqn{\chi_1^0(\tau) = \frac{\tau}{2 \Delta x} h_{\rho\sigma}(x_1)
\Delta x^{\rho} \Delta x^{\sigma} } \nonumber \\
& & + \frac{\tau}{\Delta x} \int_0^1 \! d\tau' \, (1 - \tau') \Delta x^i
\Delta \Gamma^i_{~\rho\sigma}\Bigl(x + \Delta x \tau\Bigr) \Delta x^{\rho}
\Delta x^{\sigma} \nonumber \\
& & \hspace{.5cm} - \int_0^{\tau} \! d\tau' \, (\tau - \tau') 
\Delta \Gamma^0_{~\rho\sigma}\Bigl(x_1 + \Delta x \tau\Bigr) \Delta x^{\rho}
\Delta x^{\sigma} \; . \qquad \label{DX0}
\end{eqnarray}
Ignoring ordinary matter makes the graviton geodesics identical to 
$\chi_0^{\mu}(\tau)$. Hence the time lag between the arrival of gravitational
waves and the arrival of photons or neutrinos is (to first order in the
potentials),
\begin{equation}
c \Delta t \equiv \chi_1^0(1) = \frac1{2 \Delta x} \int_0^1 \! d\tau 
\, h_{\mu\nu}\Bigl(x_1 + \Delta x \tau\Bigr) \Delta x^{\mu} 
\Delta x^{\nu} \; . \label{thelag}
\end{equation}

Expression (\ref{thelag}) can be simplified a great deal further. First
expand out the potentials,
\begin{equation}
h_{\mu\nu} \Delta x^{\mu} \Delta x^{\nu} = - \Delta B \, \Delta x^2
+ \Delta A \, (\hat{r} \cdot \Delta \vec{x})^2 \; .
\end{equation}
(Note that the time lag is positive semi-definite because $\Delta B \leq
0$ and $\Delta A \geq 0$.) Now use relation (\ref{DBfromDA}) for $\Delta 
B(r)$ in terms of $\Delta A(r)$ and partially integrate to reach the form,
\begin{eqnarray}
\lefteqn{c\Delta t = -\frac{\Delta x}{2} \Delta B(r_2) } \nonumber \\
& & \hspace{.5cm} + \frac{\Delta x}{2} \int_0^1 \! d\tau \, \Delta A(r) 
\Biggl[ \frac{\tau}{r} \frac{\partial r}{\partial \tau} + \Bigl( \frac{\hat{r} 
\cdot \Delta \vec{x}}{\Delta x}\Bigr)^2 \Biggr] \; , \qquad \\
& & = \frac{\Delta \vec{x} \cdot \vec{x}_1}{2 \Delta x} \, \Delta B(r_1)
- \frac{\Delta \vec{x} \cdot \vec{x}_2}{2 \Delta x} \, \Delta B(r_2) 
\nonumber \\
& & \hspace{.5cm} + \Delta x \int_0^1 \!d\tau \, \Delta A(r) \Biggl[1 +
\frac{ (\vec{x}_1 \cdot \Delta \vec{x})^2 - r_1^2 \Delta x^2}{\Delta x^2 r^2}
\Biggr] \; . \qquad
\end{eqnarray}
It is useful to define the constant $C$,
\begin{equation}
C \equiv \frac1{\Delta x^2} \sqrt{r_1^2 \Delta x^2 - (\vec{x}_1 \cdot
\Delta \vec{x})^2} \; .
\end{equation}
Finally, we change variables from $\tau$ to $r$,
\begin{equation}
r(\tau) = \Delta x \sqrt{\Bigl(\tau + \Bigl(\frac{\vec{x}_1 \cdot \Delta 
\vec{x}}{\Delta x^2}\Bigr)^2 + C^2} \; .
\end{equation}
Assuming $r_2 < r_1$ the result is,
\begin{eqnarray}
\lefteqn{c \Delta t = \frac{\Delta \vec{x} \cdot \vec{x}_1}{2 \Delta x} \, 
\Delta B(r_1) - \frac{\Delta \vec{x} \cdot \vec{x}_2}{2 \Delta x} \, 
\Delta B(r_2) } \nonumber \\
& & \hspace{2cm} + \int_{r_2}^{r_1} \!\!\! dr \, \frac{2 G M(r)}{c^2 r}
\sqrt{1 - \Bigl(\frac{C \Delta x}{r}\Bigr)^2 } \; . \qquad \label{cdt}
\end{eqnarray}
For $r_1 < r_2$ we take the other root of the solution for $\tau(r)$,
which reverses the upper and lower limits in (\ref{cdt}).

\section{Results}

We have worked out explicit results for three typical sources at vastly
different distances: GRB 070201, SN 1987a and Sco-X1. Their celestial
coordinates are given in Table~\ref{Coords}. Table~\ref{Coords} also 
gives the centers of the Milky Way and Andromeda dark matter halos.

\begin{table}

\vbox{\tabskip=0pt \offinterlineskip
\def\tablerule{\noalign{\hrule}}
\halign to245pt {\strut#& \vrule#\tabskip=1em plus2em& \hfil#\hfil&
\vrule#& \hfil#\hfil& 
\vrule#& \hfil#\hfil& \vrule#& \hfil#\hfil& \vrule#\tabskip=0pt\cr
\tablerule 
\omit&height4pt&\omit&&\omit&&\omit&&\omit&\cr
&&$\!\!\!\!{\rm Object}\!\!\!\!\!\!\!$ &&$\!\!\!\!\!{\rm R.\ Ascen.}
\!\!\!\!\!\!$ && $\!\!\!\!\! {\rm Decl.} \!\!\!\!\!$ && $\!\!\!\!\! {\rm Dist.}
\!\!\!\!\!$ & \cr
\omit&height4pt&\omit&&\omit&&\omit&&\omit&\cr
\tablerule
\omit&height4pt&\omit&&\omit&&\omit&&\omit&\cr
&& $\!\!\!\!\! {\rm Milky\ Way} \!\!\!\!\!\!\!$
&& $\!\!\!\!\!17{\rm h} \, 45{\rm m} \, 40{\rm s} \!\!\!\!\!\!\!$
&& $\!\!\!\!\! {\scriptstyle -}29^{\circ} \, 00' \, 28'' \!\!\!\!\!\!\!$
&& $\!\!\!\!\! 7.94~{\rm kpc} \!\!\!\!\!\!\!$ & \cr
\omit&height4pt&\omit&&\omit&&\omit&&\omit&\cr
\tablerule
\omit&height4pt&\omit&&\omit&&\omit&&\omit&\cr
&& $\!\!\!\!\! {\rm Andromeda} \!\!\!\!\!\!\!$
&& $\!\!\!\!\!00{\rm h} \, 42{\rm m} \, 44{\rm s} \!\!\!\!\!\!\!$
&& $\!\!\!\!\! {\scriptstyle +}41^{\circ} \, 16' \, 09'' \!\!\!\!\!\!\!$
&& $\!\!\!\!\! 778~{\rm kpc} \!\!\!\!\!\!\!$ & \cr
\omit&height4pt&\omit&&\omit&&\omit&&\omit&\cr
\tablerule
\omit&height4pt&\omit&&\omit&&\omit&&\omit&\cr
&& $\!\!\!\!\! {\rm GRB\ 070201} \!\!\!\!\!\!\!$
&& $\!\!\!\!\!00{\rm h} \, 44{\rm m} \, 32{\rm s} \!\!\!\!\!\!\!$
&& $\!\!\!\!\! {\scriptstyle +}42^{\circ} \, 14' \, 21'' \!\!\!\!\!\!\!$
&& $\!\!\!\!\! 780~{\rm kpc} \!\!\!\!\!\!\!$ & \cr
\omit&height4pt&\omit&&\omit&&\omit&&\omit&\cr
\tablerule
\omit&height4pt&\omit&&\omit&&\omit&&\omit&\cr
&& $\!\!\!\!\! {\rm SN\ 1987a} \!\!\!\!\!\!\!$
&& $\!\!\!\!\!05{\rm h} \, 35{\rm m} \, 28{\rm s} \!\!\!\!\!\!\!$
&& $\!\!\!\!\! {\scriptstyle -}69^{\circ} \, 16' \, 12'' \!\!\!\!\!\!\!$
&& $\!\!\!\!\! 51.4~{\rm kpc} \!\!\!\!\!\!\!$ & \cr
\omit&height4pt&\omit&&\omit&&\omit&&\omit&\cr
\tablerule
\omit&height4pt&\omit&&\omit&&\omit&&\omit&\cr
&& $\!\!\!\!\! {\rm Sco}{\scriptstyle -}{\rm X1} \!\!\!\!\!\!\!$
&& $\!\!\!\!\!16{\rm h} \, 19{\rm m} \, 55{\rm s} \!\!\!\!\!\!\!$
&& $\!\!\!\!\! {\scriptstyle -}15^{\circ} \, 38' \, 24'' \!\!\!\!\!\!\!$
&& $\!\!\!\!\! 2.80~{\rm kpc} \!\!\!\!\!\!\!$ & \cr
\omit&height4pt&\omit&&\omit&&\omit&&\omit&\cr
\tablerule}}

\caption{Angular coordinates and distances for the Milky Way and 
Andromeda galaxies and for the three sources of this study.}
\label{Coords}
\end{table}

GRB 070201 was a short hard gamma ray burst whose angular error 
box corresponded to a  $0.124^{\circ}$ quadrilateral which overlapped with the Andromeda galaxy~\cite{Mazets07}. 
Short hard gamma-ray bursts are believed to be caused by the mergers of two
neutron stars or a neutron star and a black hole~\cite{Nakar07}.
If GRB 070201 derived from such a merger, with masses close to 
1.4 $M_{\odot}$ and a reasonable orientation, the GW signal should 
have been seen if its distance  was $780~{\rm kpc}$~\cite{Anderson07}.
It is however possible that the GRB did not originate in the Andromeda 
galaxy, and in that case the signal from a compact object merger 
may be inaccessible to LIGO.

No gravitational waves were found from a search done within
$\pm$ 180~s time-window with the LIGO Hanford detectors around the 
time of this GRB~\cite{070201,Dietz08}. One interpretation of this
null result is that GRB 070201 was a SGR flare~\cite{070201,Ofek07}.
However, it is also possible that physics is described by a Dark
Matter Emulator, in which case the pulse of gravitational waves would have
arrived long before the electromagnetic signal. Table~\ref{Results}
gives our results for the time lag one would expect at the central
position using each of the three dark matter density profiles. 
Although the time lags differ by as much as 69 days, none of the
lags is less than two years. 

Table~\ref{Errors} considers another measure of the likely error by 
specializing to the isothermal profile and varying the angular 
position (at the fixed distance of $780~{\rm kpc}$) over the four 
vertices of the angular error box. In this case distinct results
are reported for the contributions from the Milky Way and Andromeda
halos, which are of course independent at linearized order. As 
expected, varying the position has no effect on the contribution 
from the Milky Way halo but it can change the contribution from the
Andromeda halo by as much as 15 days. We should however stress that if this delay calculation
was done by assuming that this GRB is at a distance of $780~{\rm kpc}$. If  
this GRB did not originate in Andromeda, then the calculated delay would be much larger.

\begin{table}

\vbox{\tabskip=0pt \offinterlineskip
\def\tablerule{\noalign{\hrule}}
\halign to245pt {\strut#& \vrule#\tabskip=1em plus2em& \hfil#\hfil&
\vrule#& \hfil#\hfil& 
\vrule#& \hfil#\hfil& \vrule#& \hfil#\hfil& \vrule#\tabskip=0pt\cr
\tablerule 
\omit&height4pt&\omit&&\omit&&\omit&&\omit&\cr
&&$\!\!\!\!{\rm Profile}\!\!\!\!\!\!\!$ &&$\!\!\!\!\!{\rm GRB\ 070201}
\!\!\!\!\!\!$ && $\!\!\!\!\! {\rm SN\ 1987a} \!\!\!\!\!$ && $\!\!\!\!\! 
{\rm Sco}{\scriptstyle -}{\rm X1} \!\!\!\!\!$ & \cr
\omit&height4pt&\omit&&\omit&&\omit&&\omit&\cr
\tablerule
\omit&height4pt&\omit&&\omit&&\omit&&\omit&\cr
&& $\!\!\!\!\! {\rm Isothermal} \!\!\!\!\!\!\!$
&& $\!\!\!\!\! 742~{\rm days} \!\!\!\!\!\!\!$
&& $\!\!\!\!\! 78.2~{\rm days} \!\!\!\!\!\!\!$
&& $\!\!\!\!\! 4.98~{\rm days} \!\!\!\!\!\!\!$ & \cr
\omit&height4pt&\omit&&\omit&&\omit&&\omit&\cr
\tablerule
\omit&height4pt&\omit&&\omit&&\omit&&\omit&\cr
&& $\!\!\!\!\! {\rm NFW} \!\!\!\!\!\!\!$
&& $\!\!\!\!\! 804~{\rm days} \!\!\!\!\!\!\!$
&& $\!\!\!\!\! 74.8~{\rm days} \!\!\!\!\!\!\!$
&& $\!\!\!\!\! 4.88~{\rm days} \!\!\!\!\!\!\!$ & \cr
\omit&height4pt&\omit&&\omit&&\omit&&\omit&\cr
\tablerule
\omit&height4pt&\omit&&\omit&&\omit&&\omit&\cr
&& $\!\!\!\!\! {\rm Moore} \!\!\!\!\!\!\!$
&& $\!\!\!\!\! 811~{\rm days} \!\!\!\!\!\!\!$
&& $\!\!\!\!\! 74.5~{\rm days} \!\!\!\!\!\!\!$
&& $\!\!\!\!\! 4.97~{\rm days} \!\!\!\!\!\!\!$ & \cr
\omit&height4pt&\omit&&\omit&&\omit&&\omit&\cr
\tablerule}}

\caption{Time delays from three dark matter profiles for
each of the three sources of this study.}
\label{Results}
\end{table}

SN 1987a was a core collapse supernova in the Large Magellanic Cloud
at a distance of 51.4~kpc~\cite{SN1987a}. Neutrinos were observed by the 
Kami\-o\-kan\-de-II~\cite{Kam1,Kam2} and 
Irvine-Michigan-Brookhaven~\cite{Bion,Brat} detectors. The optical 
signal arrived several hours later because photons must traverse the
optically dense stellar environment~\cite{SN1987a}. The total Shapiro delay 
for SN 1987A from the visible and dark matter distribution 
has also been independently estimated~\cite{Longo88,Mena07} to be between 0.29 to 0.36 years.
If the oblateness 
of SN 1987a was in relation to that of the Sun, the current gravitational 
wave detectors would probably not have seen anything had they been
operating at the time~\cite{Dimmel}. However, advanced LIGO would detect 
such a supernova out to $0.8~{\rm Mpc}$~\cite{Dimmel}. This includes the 
Andromeda galaxy, which doubles the expected rate and also ensures that 
the signal passes through dark matter dominated regions. Of course the 
effective coverage from neutrino detectors will remain limited to our 
galaxy and its satellites~\cite{BFV,Vogel}.

Table~\ref{Results} gives our results for the expected time lag from a
Dark Matter Emulator which reproduces each of the three dark matter 
profiles. These results include only the effect of the Milky 
Way halo. In contrast to the much more distant GRB 070201, the scatter 
between the various models for SN 1987a is much smaller --- a mere 2.7 days.

Sco-X1 (located at a distance of 2.8 kpc) is one of the brightest Low
Mass X-ray Binaries (LMXBs). LMXBs are potential sources of gravitational 
waves from r-modes getting excited due to accretion, or from a deformed
crust~\cite{Bildsten98,Andersson99,Heyl02}. One proposed search is to look
for coincidences between the data from LIGO and Rossi X-Ray Timing
satellite~\cite{extrig08}. This search also assumes that gravitational
waves and X-ray photons arrive at the same time.

Table~\ref{Results} reports the expected time lag for each of the three 
dark matter profiles, again from the Milky Way halo. Although the time
lag is still easily observable at $\sim 4.9~{\rm days}$, the agreement
between the three models for this source is excellent. The largest 
discrepancy is just $.1~{\rm day}$.

\begin{table}

\vbox{\tabskip=0pt \offinterlineskip
\def\tablerule{\noalign{\hrule}}
\halign to245pt {\strut#& \vrule#\tabskip=1em plus2em& \hfil#\hfil&
\vrule#& \hfil#\hfil& 
\vrule#& \hfil#\hfil& \vrule#& \hfil#\hfil& \vrule#\tabskip=0pt\cr
\tablerule 
\omit&height4pt&\omit&&\omit&&\omit&&\omit&\cr
&&$\!\!\!\!{\rm R.\ Ascension}\!\!\!\!$ && $\!\!\!\!\! {\rm Declination}
\!\!\!\!\!$ && $\!\!\!\!\! \Delta t_{\rm MW} \!\!\!\!\!$ && $\!\!\!\!\! 
\Delta t_{\rm M31} \!\!\!\!\!$ & \cr 
\omit&height4pt&\omit&&\omit&&\omit&&\omit&\cr
\tablerule 
\omit&height4pt&\omit&&\omit&&\omit&&\omit&\cr
&& $\!\!\!\!00{\rm h} \, 44{\rm m} \, 32{\rm s} \!\!\!\!$ 
&& $\!\!\!\!\! 42^{\circ} \, 14' \, 21'' \!\!\!\!\!$ 
&& $\!\!\!\!\! 407~{\rm dy} \!\!\!\!\!$ 
&& $\!\!\!\!\! 335~{\rm dy} \!\!\!\!\!$ & \cr 
\omit&height4pt&\omit&&\omit&&\omit&&\omit&\cr
\tablerule 
\omit&height4pt&\omit&&\omit&&\omit&&\omit&\cr
&& $\!\!\!\!00{\rm h} \, 46{\rm m} \, 18{\rm s} \!\!\!\!$ 
&& $\!\!\!\!\! 41^{\circ} \, 56' \, 42'' \!\!\!\!\!$ 
&& $\!\!\!\!\! 407~{\rm dy} \!\!\!\!\!$ 
&& $\!\!\!\!\! 337~{\rm dy} \!\!\!\!\!$ & \cr 
\omit&height4pt&\omit&&\omit&&\omit&&\omit&\cr
\tablerule 
\omit&height4pt&\omit&&\omit&&\omit&&\omit&\cr
&& $\!\!\!\!00{\rm h} \, 41{\rm m} \, 51{\rm s} \!\!\!\!$ 
&& $\!\!\!\!\! 42^{\circ} \, 52' \, 08'' \!\!\!\!\!$ 
&& $\!\!\!\!\! 407~{\rm dy} \!\!\!\!\!$ 
&& $\!\!\!\!\! 322~{\rm dy} \!\!\!\!\!$ & \cr 
\omit&height4pt&\omit&&\omit&&\omit&&\omit&\cr
\tablerule 
\omit&height4pt&\omit&&\omit&&\omit&&\omit&\cr
&& $\!\!\!\!00{\rm h} \, 42{\rm m} \, 47{\rm s} \!\!\!\!$ 
&& $\!\!\!\!\! 42^{\circ} \, 31' \, 41'' \!\!\!\!\!$ 
&& $\!\!\!\!\! 407~{\rm dy} \!\!\!\!\!$ 
&& $\!\!\!\!\! 330~{\rm dy} \!\!\!\!\!$ & \cr 
\omit&height4pt&\omit&&\omit&&\omit&&\omit&\cr
\tablerule 
\omit&height4pt&\omit&&\omit&&\omit&&\omit&\cr
&& $\!\!\!\!00{\rm h} \, 47{\rm m} \, 14{\rm s} \!\!\!\!$ 
&& $\!\!\!\!\! 41^{\circ} \, 35' \, 35'' \!\!\!\!\!$ 
&& $\!\!\!\!\! 407~{\rm dy} \!\!\!\!\!$ 
&& $\!\!\!\!\! 338~{\rm dy} \!\!\!\!\!$ & \cr 
\omit&height4pt&\omit&&\omit&&\omit&&\omit&\cr
\tablerule}}

\caption{Shapiro Delays for GRB 070201 from the Isothermal 
Profiles of the Milky Way ($\Delta t_{\rm MW}$) and Andromeda 
($\Delta t_{\rm M31}$) at the central value of the angular 
position and at the four vertices of the error box. In all 
cases the distance to the burst was taken to be $780~{\rm kpc}$.}
\label{Errors}
\end{table}

\section{Other Modified Gravity Theories}

Since the 1970s, there have been various proposed tests of general 
relativity through gravitational wave observations (See Ref.~\cite{Alexander07}
for a recent review). Most of these tests are in the strong field regime. 
In this section, we list some other non-GR gravity theories which also 
predict a non-zero time-delay between photons and GWs and are not yet ruled 
out through other observations.

These are massive graviton theories and brane-world models. In massive 
graviton theories~\cite{Will97}, the gravitational waves would arrive {\it 
after} the photons, with the delay being dependent on the graviton mass. 
However, the Moore-Nelson lower bound on the speed of gravitational waves 
imposes stringent constraints on the validity of massive graviton models. 
It should be noted that there is no consistent interacting theory for 
massive spin two particles which is limited to a finite number of 
fields~\cite{Deser}.

In various brane-world models, gravitational waves propagate faster than 
photons or neutrinos, depending on the curvature of the 
bulk~\cite{Csaki01,Chung02,Ahmadi07}. Therefore, it is not possible to 
calculate model-independent time-delays for the three sources we considered 
in this paper.

\section{Conclusions}

The power and generality of our analysis derives from ignoring the 
details of how a Dark Matter Emulator dispenses with the need for 
galactic dark matter. We merely assume that it does, which implies that 
ordinary matter must couple to the metric that general relativity would 
predict with dark matter. The special characteristic of Dark Matter 
Emulators is that weak gravitational waves couple to the metric that general 
relativity would predict without dark matter. {\it Both of these metrics 
can be inferred from observation, and all geometrical quantities worked 
out, without regard to the details of specific models.} Although a Dark
Matter Emulator is not the only conceivable way of evading the no-go 
theorem~\cite{SW2} while preserving solar system tests~\cite{Will}, 
it is the only way that has so far been given a concrete realization.

If dark matter does not exist and the observed cosmic motions and
lensing instead derive from a Dark Matter Emulator then the assumption
upon which all triggered gravitational wave searches are based breaks down. 
In this case the optical or neutrino identification of a plausible 
gravitational wave source would not imply a simultaneous pulse of gravitational
waves but rather that such a pulse occurred {\it earlier}. Even for 
nearby sources such as Sco-X1 (at about $2.8~{\rm kpc}$) the gravitational
waves would arrive almost five days earlier. For a source in the 
Andromeda galaxy the time difference would be over two years.

It is obviously premature to proclaim that the failure of triggered
searches to reveal any coincident gravitational wave pulse implies that
physics is described by a Dark Matter Emulator. But if plausible
sources continue to produce null results this possibility has to be 
considered. In that case the key question becomes the accuracy with
which one can estimate the expected time lag. Some measure of this
is given by the spread in Table~\ref{Results} for different reasonable
dark matter profiles. Table~\ref{Errors} considers variations in the
angular position, and there will be comparable results for varying the
much less well-determined distances. Based on these analyses it seems
unlikely that the uncertainty can be reduced below the level of a few 
percent. This has important implications for the way data needs to be 
kept and for the types of searches that should be contemplated.

Of course a gravitational wave signal might be loud enough to show up in
all-sky untriggered searches~\cite{LIGOs4}. In that case the gravitational
wave signal would trigger electromagnetic and neutrino searches. A 
single gold-plated event in which the counterpart signal arrived after
a plausible delay would be powerful evidence in support of Dark Matter 
Emulators. Conversely, a single detection of coincident signals would
rule out the entire class of Dark Matter Emulators. This is a novel way
of using gravitational wave detectors to test alternate gravity theories in
the ultra-weak field regime. Indeed, this is an ideal test because Dark 
Matter Emulators do not change aspects of the tensor component of a 
gravitational wave signal such as the number of polarizations or, to 
any reasonable accuracy, the travel time between Earth-bound detectors. 
So there need be no change in the data analysis algorithms that would 
be used in any case.

\begin{acknowledgments}
We would like to thank H.~Asada, P.~Gondolo, J.~Kanner, B.~Kocsis, G.~Lake, 
S.~Marka, G.~Moore, E.~Tempel and B.~Whiting. This work was partially 
supported by NSF grants NSF-428-51 29NV0 and PHY-0653085, by the Institute for Fundamental Theory 
at the University of Florida and by the Center for Gravitational Wave Physics 
at Penn State. The Center for Gravitational Wave Physics is funded by the 
National Science Foundation under Cooperative Agreement PHY 01-14375.
\end{acknowledgments}

\end{document}